\documentclass[bimj,fleqn]{w-art-preprint}
\usepackage{times}
\usepackage{w-thm}
\usepackage[authoryear]{natbib}
\theoremstyle{plain}

\theoremstyle{definition}

\usepackage[]{graphicx}
\chardef\bslash=`\\ 

\usepackage{xurl} 
\usepackage{amssymb}
\usepackage{hyperref}
\hypersetup{hidelinks,breaklinks=true}
\usepackage{float} 
\usepackage{xurl}
\makeatletter
\providecommand{\@bibauthor}[1]{#1}
\providecommand{\@bibjournal}[1]{#1}
\providecommand{\@bib@Z@journal}[1]{#1}
\providecommand{\@bibibid@}[1]{#1}
\makeatother





\begin{document}
\pagespan{1}{}
\keywords{Stochastic EM Algorithm, Skellam distribution, Survival, COVID-19, ICU-patients\\
\noindent\hspace*{-4.2pc} The GitHub repository accompanying this paper can be found under\break \hspace*{-4pc} \url{https://github.com/MartjeRave/OccupancyDuration.git}} 
\title[DurationTimeOccupancyData]{Deriving Duration Time from Occupancy Data – A case study in the length of stay in Intensive Care Units for COVID-19 patients}
\author[Martje Rave]{Martje Rave\footnote{Corresponding author: {\sf{e-mail: martje.rave@stat.uni-muenchen.de}}, Phone: (+49)-89-2180-2248, Fax: (+49)-89-2180-5040}\inst{1}} 
\address[\inst{1}]{Chair of Applied Statistics in Social Sciences, Economics and Business, 
             Department of Statistics,
             Faculty of Mathematics, Informatics and Statistics,
             Ludwig-Maximilians-Universität München,
             Ludwigstr. 33,
             80539 München,
             Germany}
\address[\inst{2}]{Munich Center for Machine Learning (MCML)}
\author[Göran Kauermann]{Göran Kauermann\inst{1, 2}}
\Receiveddate{} \Reviseddate{} \Accepteddate{} 

\begin{abstract}
This paper focuses on drawing information on underlying processes, which are not directly observed in the data. In particular, we work with data in which only the total count of units in a system at a given time point is observed, but the underlying process of inflows, length of stay and outflows is not. The particular data example looked at in this paper is the occupancy of intensive care units (ICU) during the COVID-19 pandemic, where the aggregated numbers of occupied beds in ICUs on the district level (`Landkreis') are recorded, but not the number of incoming and outgoing patients. The Skellam distribution allows us to infer the number of incoming and outgoing patients from the occupancy in the ICUs. This paper goes a step beyond and approaches the question of whether we can also estimate the average length of stay of ICU patients. Hence, the task is to derive not only the number of incoming and outgoing units from a total net count but also to gain information on the duration time of patients on ICUs. We make use of a stochastic Expectation-Maximisation algorithm and additionally include exogenous information which are assumed to explain the intensity of inflow.
\end{abstract}

\maketitle

\section{Introduction}\label{sec:Intro}

In this paper, we introduce a method which enables one to estimate an underlying, unobserved inflow, length of stay, and consequent outflow of units, using only sporadically observed net count data of said units. 
While we look at intensive care units (ICU) in the paper, we want to make clear right at the start that the approach is transferable to similar data constellations. Consider, for instance, the research of an ornithologist who is investigating the hatches and deaths in a given penguin colony. The scientist sporadically observes the number of penguins at given time points. Between each observation, some penguins will have hatched and some will have died. The methodology developed in this paper allows us to estimate the number of incoming units (hatched penguins), the  length of stay (average life span) and the number of outgoing units (penguins which have died). 
The same question is posed on our data example. We observe data on the occupancy of ICUs during the COVID-19 pandemic, but the real focus of interest is on obtaining information of incoming and outgoing patients as well as on the (average) length of stay in the ICU. 

Throughout the COVID-19 pandemic, there were arguably a good amount of data published in Germany, foremost by the Robert Koch Institute (RKI), on COVID-19 infections, and the Deutsche Interdisziplinäre Vereinigung für Intensiv- und Notfallmedizin (DIVI), on hospital and ICU occupancy. However, in the beginning of the pandemic there were no data published on the number of incoming and outgoing ICU patients infected with COVID-19, only the ICU occupancy. While these data were made available on the state level (`Bundeslandebene') from 2021 onwards, data on district level- which is what we consider in this paper- have not been published. 

Data on ICU admissions for the whole of Germany were analysed by, for example, \citep{DifferenceICUWave} to evidence the difference in the initial pandemic waves. Others, particularly medical practitioners, conducted studies on individual treatment centre level, to assess the treatment strategy and the success thereof, see e.g. \citep{FreiburgICU}. 

In our earlier work, \citep{OccupancyData}, we analyse the occupancy in relation to the infection rates in order to understand the strain on the healthcare system.
Clearly, the occupancy is a function of admission and length of stay. This is the core assumption in our subsequent work \citep{rave2023skellam}, in which we take the length of stay as fixed, relying on results of \citep{Tolksdorf2020}.
Here, we extend our previous work and demonstrate, that the length of stay can also be estimated from occupancy data, besides obtaining information on incoming and outgoing patients. By doing so, our approach allows us to gain more understanding of the epidemiological dynamics of the disease.

The key component of our statistical model looks at the difference in two independent counting processes, each assumed to be Poisson distributed. This leads to a Skellam distribution \citep{skellam1948probability} with parameters equivalent to the intensity parameters of the respective underlying in- and outgoing Poisson processes.  We model the unobserved number of incoming and outgoing units to depend on a set of covariates, as well as spatio-temporal information. The required independence of the two Poisson processes is achieved by conditioning on the history of the process, i.e., we assume some Markov structure. 


Fitting is pursued by applying the stochastic Expectation-Maximisation (sEM) algorithm as introduced by \citep{celeux1996stochastic} and further discussed among others in \citep{NEURIPS2018_aba22f74} concerning running time or \citep{sEMCovidLatent2023} for estimation of complex or uncommon distributions; see also \citep{yang2016stochastic} for latent variable estimation in survival models.
In our application, we iteratively and sequentially simulate the number of incoming and outgoing units, using the Skellam distribution. This replaces the unobserved values by simulated values (stochastic E step), and the sequential simulation allows us to condition on the past, so that we can utilise the Markov structure in the simulations. The E-step provides a complete data set which is used to estimate the incoming and outgoing intensity parameter (M-Step) employing two independent Generalised Additive Models (GAMs), \citep{WOOD:2017}. The outgoing intensity is modelled to depend on the (unobserved) number of incoming patients, which allows to model the average length of stay of COVID-19 patients on ICUs. This part of the model is non-standard and requires specially tailored estimation routines. In simulations, we demonstrate the usability of our estimates and apply the routine to real data, as described above. 

The paper is organised as follows. In Section \ref{sec:Data}, we describe the COVID-19 ICU data in more detail. In Section \ref{sec:InOut}, we go into further detail of our estimation process, by describing the sEM algorithm, first through our initial approach, then by our extension thereof. We then show the application to simulated data in Section \ref{sec:SimApp} and the application to COVID-19 ICU data in Section \ref{sec:CovApp}. We discuss the method in Section \ref{sec:Diss}. 

\section{COVID-19 ICU Data}\label{sec:Data}
We define with $Y_{(t,d)}$ the observed COVID-19-related occupancy of the ICUs at time point $t$ in the administrative district $d$. For time, we take the interval   $1^{st}$ of August 2021 to the $31^{st}$ of December 2021 with $t=1,2, \ldots $ denoting the days. For the districts, we have a total of $400$ different administrative regions, districts, in Germany. Data on the ICU occupancy are provided by DIVI\footnote[1]{\href{https://robert-koch-institut.github.io/Intensivkapazitaeten_und_COVID-19-Intensivbettenbelegung_in_Deutschland/}{https://robert-koch-institut.github.io/Intensivkapazitaeten\_und\_COVID-19-Intensivbettenbelegung\_in\_Deutschland/}} \citep{DIVI_REPO}, and additionally, we take the daily infection rates as provided by the RKI\footnote[2]{\href{https://robert-koch-institut.github.io/COVID-19_7-Tage-Inzidenz_in_Deutschland/}{https://robert-koch-institut.github.io/COVID-19\_7-Tage-Inzidenz\_in\_Deutschland/}} \citep{RKI_Inzidenz_REPO}. 

To the best of our knowledge, there are no data on the incoming, length of stay or mortality of ICU patients infected with COVID-19, publicly available in Germany. So in order to later link the inflow and outflow of patients to data, which are observed, we take the ICU occupancy and we can calculate its difference 
\begin{equation}\label{eq:OccDiff}
\Delta_{(t,d)}=Y_{(t,d)}-Y_{(t-1,d)}. 
\end{equation}

The infection rates are provided for each district, day and age group, namely `35-59' year olds, `60-79' year olds and `80+' year olds. We calculate the 7-day-average of the infection rate per $100.000$ inhabitants and take the natural logarithm thereof.

\begin{figure}
    \centering
    \includegraphics[width=\linewidth]{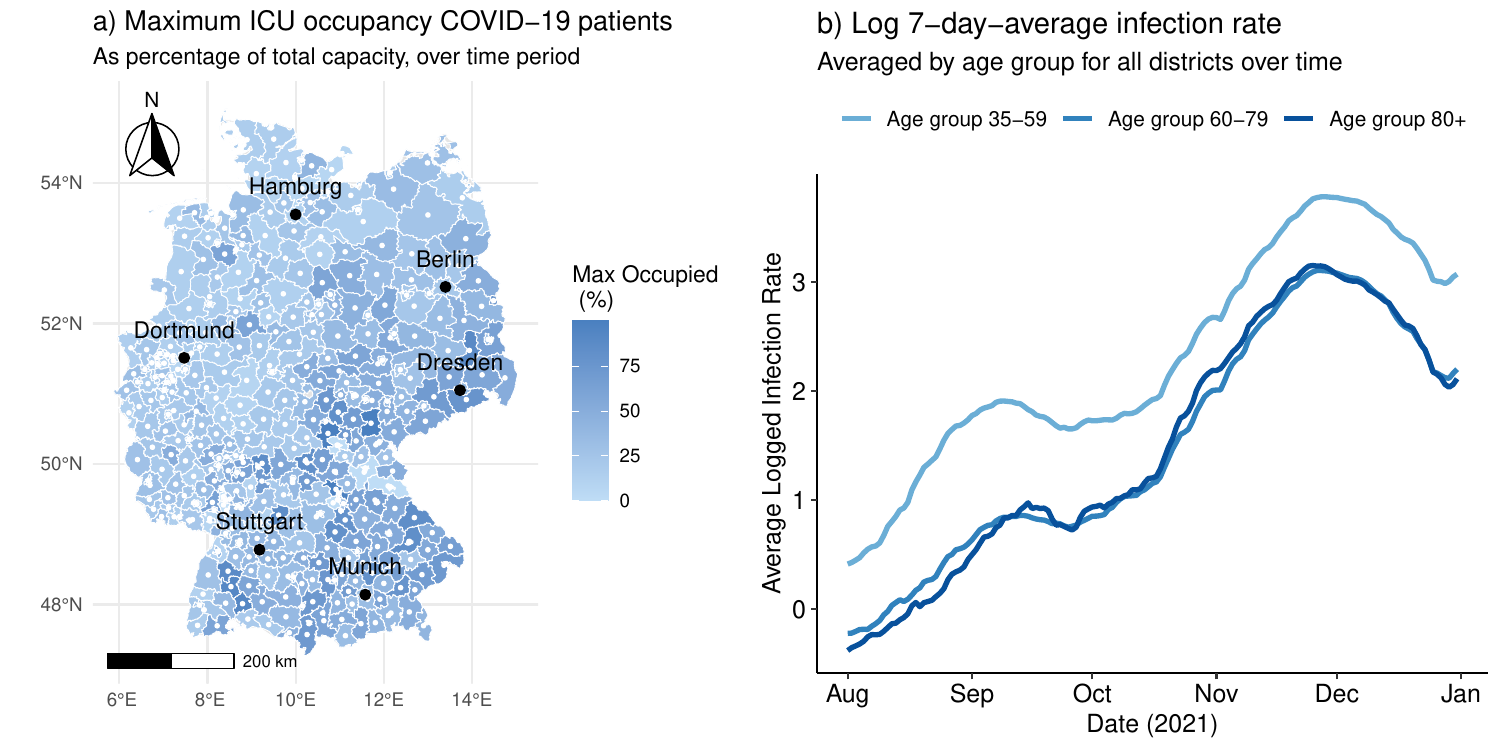}
    \caption{Introduction to COVID-19 Data with; a) maximum ICU occupancy, as a percentage of the total ICU beds, per district (`Landkreis') and b) average of the logged 7-day-average infection rate per 100.000 inhabitants, per age group from the $1^{st}$ of August until the $31^{st}$ of December, 2021.}
    \label{fig:IntroductionData}
\end{figure}

For data exploration, we plot the maximum ICU occupancy in Figure \ref{fig:IntroductionData} a). We show the percentage COVID-19 occupation of the total ICU beds, per district.  The logged 7-day-average infection rates per 100.000 inhabitants of the three age groups included in the analysis, plotted over time and averaged for all districts in Germany are visualised in Figure \ref{fig:IntroductionData} b). 

Figure \ref{fig:IntroductionData} a) shows a somewhat constant maximum occupancy rate over space, with some rural districts exhibiting a larger occupancy rate than cities. One might add that some districts report to have as little as $6$ ICU beds available for patients. We would therefore expect to see these filled up more quickly than others. The cities Hamburg and Berlin are observed to have a maximum occupancy of COVID-19 patients of around $12.6\%$ and $8.2\%$, respectively. Dresden, München and Stuttgart are observed to have a maximum occupancy of around $22\%$ to $26\%$. Dortmund's maximum occupancy is observed at around $38\%$. More information on the ICU dynamics in Germany are published by the \citep{Gesundheit_2025}. The centroids of the given districts are marked by the respective white dots seen in Figure \ref{fig:IntroductionData} a).

Figure \ref{fig:IntroductionData} b) shows two spikes in the average of the logged infection rate per 100.000 inhabitants, per age group; one in mid September and another, larger spike, at the end of November, 2021. While there were non-pharmaceutical interventions in place, such as curfews and testing strategies, some readers might remember a dramatic infection wave towards the end of the second half of 2021. We also observe this in Figure \ref{fig:IntroductionData} b). 

\section{Modelling Incoming and Outgoing}\label{sec:InOut}
\subsection{Skellam Modell}\label{sec:Skellam}

We are interested in the underlying process of incoming, length of stay and outgoing units, which is not observed. We therefore define with $I_{(t,d)}$ the incoming and with  $R_{(t,d)}$ the outgoing (released) units of the ICUs in district $d$ at time point $t$.  We use the equivalence between $\Delta_{(t,d)}$, as given in (\ref{eq:OccDiff}), and the difference between the incoming units and outgoing units, i.e.
\begin{align}
\label{eq:Delta}
\Delta_{(t,d)}=Y_{(t,d)}-Y_{(t-1,d)}=I_{(t,d)}-R_{(t,d)}.
\end{align}

As $I_{(t,d)}$ and $R_{(t,d)}$ are both counting processes, it is reasonable to assume that the two random variables follow Poisson distributions, with intensity parameters $\lambda^I_{(t,d)}$ and $\lambda^R_{(t,d)}$, respectively, i.e.

\begin{align}
\label{eq:in}
I_{(t,d)}  & \sim \mbox{Poisson}\left( \lambda^I_{(t,d)}\right) \\
\label{eq:out}
R_{(t,d)}  & \sim \mbox{Poisson}\left( \lambda^R_{(t,d)} \right).
\end{align} 
We define with $H_{t,d}$ the history of the incoming process, that is $H_{t,d} = \{ I_{\tilde t} : \tilde{t} < t\}$. Given the history of the incoming, we assume that $I_{t,d}$ and $R_{t,d}$ are conditionally independent, so that the difference of the two Poisson distributions follows the Skellam distribution, \citep{skellam1948probability},
\begin{align}\label{eq:DiffEq}
\Delta_{(t,d)} | H_{t,d} \sim \mbox{Skellam}(\lambda^I_{(t,d)}, \lambda^R_{(t,d)}).
\end{align}

For the incoming intensity we set 
\begin{align}
\label{eq:model-in-est}
\lambda^I_{(t,d)}& = \exp\left( \eta_{(t,d)}^I \right)
\end{align}
where the linear predictor $\eta_{(t,d)}^I $ is modelled to depend on explanatory variables denoted by $\boldsymbol{x}^{I}_{(t,d)}$. 

The linear predictor in estimating the number of incoming ICU patients with COVID-19 is defined as 

\begin{align}\label{eq:IncEstCov}
{\lambda^I}_{(t,d)}&=\exp ({\beta_0}+{\beta_1}{\mbox{Infec}_{35-59}}_{(t,d)}+{\beta_2}{\mbox{Infec}_{60-79}}_{(t,d)}+\\
&\nonumber {\beta_3}{\mbox{Infec}_{80}}_{(t,d)}+
{\beta_4}\mbox{Monday}_{(t,d)}+\\
&\nonumber {\beta_5}\mbox{Tuesday}_{(t,d)}+ 
{\beta_6}\mbox{Wednesday}_{(t,d)}+ \\
&\nonumber {\beta_7}\mbox{Thursday}_{(t,d)}+ 
{\beta_8}\mbox{Saturday}_{(t, d)}+ 
{\beta_9}\mbox{Sunday}_{(t, d)}+\\
&\nonumber {f_1}(\mbox{long}_{(d)}, \mbox{lat}_{(d)})+{f_2}(t)) .
\end{align}
The variables included are the logged 7-day-average infection rate of the week prior to $t$ for the age groups, `35-59' year olds, `60-79' year olds and `80+' year olds. We further include a weekday effect through a dummy-coded categorical variable, with `Monday', `Tuesday', `Wednesday', `Thursday', `Saturday', `Sunday' denoting dummy indicator variables for respective weekdays and `Friday' being the reference category. Information on space is included by ${f_1}(\mbox{long}_{( d)}, \mbox{lat}_{(d)})$, a thin plate spline over the longitude and latitude of the districts' centroids. The function ${f_2}(t)$ denotes a thin plate spline across the date of observation, $t$.

For the outgoing units, $R_{(t,d)}$, we come to the understanding that this number depends on the count of incoming patients.
This is modelled multiplicative as follows. 
Let $l$ denote the time delay, i.e. the time between admission to the ICU and the current day $t$. We define with parameters $\omega_{l}$ for $l=1, \ldots , L$ the exit rates, comparable to the hazard of leaving the ICU, where $L$ is the maximum length of stay which is taken sufficiently large. 
One may also take the intensity of the number of outgoing units to be a function of external information, defined by a linear predictor $\eta_{(t,d)}^R$ which can depend on covariates $\boldsymbol{x}^{R}_{(t,d)}$. This leads to the model 
\begin{align}
\label{eq:model-out}
\lambda^R_{(t,d)} & = \exp \{ \eta^R_{(t,d)} + \log( \sum^{L}_{l=1}\omega_{l} I_{(t-l,d)})\} .
\end{align}
In our example we will simplify the setup and set $\eta^R_{(t,d)}\equiv0$. Moreover, as argued before,   $\lambda^R_{(t,d)}$ is assumed to be a function of the incoming units and the length of stay. We thus need to postulate constraints on the parameters $\omega_l$, namely 
\begin{equation}
    \label{eq:const}
\sum_{l=1}^L\omega_{l} = 1 
 \mbox{ with } \omega_{l} \geq 0~~\forall~ l~ \in \{1,\dots, L\}.
\end{equation}
for a sufficiently large $L$.

Assuming  $I_{(t,d)}$ and $R_{(t,d)}$ to be known, the estimation of the parameters of $\lambda^I_{(t,d)}$ and $\lambda^R_{(t,d)}$ would be conceptionally simple. Following the distributional assumption of (\ref{eq:in}), we would be able to maximise the likelihood, given the incoming intensity parameter using a Generalized Linear Model \citep{WOOD:2017}. The maximization of the likelihood of the outgoing number of units is, however, a little more intricate. We again assume a Poisson distribution leading to the (partial) log-likelihood
\begin{align}
    \l^R _P(\boldsymbol{\omega}) = \sum _{t=1}^T \sum _{d=1}^D\left( R_{(t,d)}\log(\sum_{l=1}^{L} \omega_l {I}_{(t-l,d)})-\sum_{l=1}^{L} \omega_l I_{(t-l,d)}\right).
    \label{eq:partlike}
\end{align}
Maximization of the log-likelihood in (\ref{eq:partlike}) needs to be done under linear constraints (\ref{eq:const}). This can be done iteratively through quadratic optimisation, see e.g.\ \citep{goldfarb_numerically_1983}.
Second-order approximation yields
\begin{align}
\label{eq:partiallike}
 l^R _P(\boldsymbol{\omega}) &\approx l^R _P(\hat{\boldsymbol{\omega}}^{(k)})+s^T(\hat{\boldsymbol{\omega}}^{(k)})(\boldsymbol{\omega}-\hat{\boldsymbol{\omega}}^{(k)})-\frac{1}{2}(\boldsymbol{\omega}-\hat{\boldsymbol{\omega}}^{(k)})^T\mathcal{I}(\hat{\boldsymbol{\omega}}^{(k)})(\boldsymbol{\omega}-\hat{\boldsymbol{\omega}}^{(k)}) \\
 \nonumber & \approx [s^T(\hat{\boldsymbol{\omega}}^{(k)})+
{\hat{\boldsymbol{\omega}}^{(k)}}{}^T \mathcal{I}(\hat{\boldsymbol{\omega}}^{(k)})]\boldsymbol{\omega}-\frac{1}{2}(\boldsymbol{\omega}^T \mathcal{I}(\hat{\boldsymbol{\omega}}^{(k)})\boldsymbol{\omega}) + K,
\end{align}
with $\hat{\boldsymbol{\omega}}^{(k)}$ denoting the estimate for the length of stay at the $k^{th}$ iteration. Quadratic optimization allows to maximize (\ref{eq:partiallike}) with respect to the linear constraints given above. More details are provided in Appendix \ref{app:scorefisher}. 

\subsection{Estimation Approach}\label{sec:EstApproach}

Since the number of incoming units and outgoing units are not observed (or observable), we can not directly estimate both, the incoming intensity (\ref{eq:model-in-est}) and outgoing intensity 
 (\ref{eq:model-out}), respectively. We therefore pursue a sEM-algorithm, where the E-step is replaced by a simulation step to iteratively obtain simulations of incoming, $\boldsymbol{I}^{(k)}$, and outgoing, $\boldsymbol{R}^{(k)}$, at the $k$-th iteration. We then use the procedures outlined in the previous subsection to estimate $\boldsymbol{\hat{\lambda}}^{I(k+1)}$ and $\boldsymbol{\hat{\lambda}}^{R(k+1)}$, given the simulated incoming and outgoing units. To be more specific, we set the parameters to some (reasonable) starting values and then simulate incoming and outgoing patients, which builds the stochastic E-step (see \citealp{celeux1996stochastic}). This leads to a complete dataset, which easily allows for (re-) estimating the parameters following the results from above. This, in turn, gives the M-step. Formally, the algorithm proceeds as follows. 

\begin{enumerate}
    \item Simulation E-Step \\
Naturally, the first observation for all districts $d=1,\dots, D$ is at $t=1$. However, since we assume $\hat{\lambda}^{R~(k)}_{(t=u,d)}=\sum_{{l}=1}^L\hat{\omega}^{(k)}_{l}{I_{(t=u-{l},d)}^{(k)}}$, for all $u=\{1,\dots, L\}$ we need the number of incoming patients before the first day of the observation period. We thus simulate $I_{(\tilde{t}, d)}^{(k)} \sim \mbox{Poisson}({\hat{\lambda}^{I~(k)}_{(t=1,d)}})$ as `burn-in', for $\tilde{t}=\{-L+1, \dots, 0\}$. 
These `burn-in' values are utilised in the E-Step simulations but not used for estimation of the incoming intensity. 
For $t=1, \dots, T$ we proceed to simulate both incoming and outgoing units conditional on the observed values $\Delta_{(t,d)}$. To be specific, we assume 
\begin{align}
I_{(t,d)}^{(k)}&\sim \mbox{Poisson} (\hat{\lambda}^{I~(k)}_{(t,d)}) \label{eq:incom_est}\\
R_{(t,d)}^{(k)}&\sim \mbox{Poisson} (\exp(\log(\sum_{l=1}^L{\hat{\omega}_{l}}^{(k)}{I^{(k)}_{(t-l,d)}}))), \label{eq:outg_est}
\end{align}
subject to
$$
I_{(t,d)}^{(k)}-R_{(t,d)}^{(k)}={Y_{(t,d)}}-{Y_{(t-1,d)}}=\Delta_{(t,d)}.
$$

Note that $I_{(t,d)}^{(k)}$ and $R_{(t,d)}^{(k)}$ are dependent and can be simulated as shown in \citep{rave2023skellam}. We reiterate the general idea, ignoring for the moment the iteration index $k$. First, we define a reasonable range $[0, I_{max}]$ of probable income values $I_{(t,d)}$. Then we calculate the truncated conditional probability 
   \begin{align}
   \label{eq:truncPMF}
    p(I_{(t,d)}=i,  ~&R_{(t,d)}=i-\Delta_{(t,d)}| I_{(t,d)} \le I_{max}; \lambda^I_{(t,d)}, \lambda^R_{(t,d)})=\\
    \nonumber
    &\frac{\exp(-\lambda^I_{(t,d)}){[\lambda^I_{(t,d)}]}^{i}~ \exp(-\lambda^R_{(t,d)})[\lambda^R_{(t,d)}]^{i-\Delta_{(t,d)}} (i! (i-\Delta_{(t,d)})!)^{-1}}{\sum_{j= 0}^{I_{max}} [\exp(-\lambda^I_{(t,d)}){[\lambda^I_{(t,d)}]}^j\exp(-\lambda^R_{(t,d)}){[\lambda^R_{(t,d)}]}^{j-\Delta_{(t,d)}} (j! (j-\Delta_{(t,d)})!)^{-1}]}.
        \end{align} 
The derivation is given in the Appendix \ref{app:jPMF}.
We then sample from this normalised truncated joint probability mass function to obtain $I_{(t,d)}^{(k)}$ and $R_{(t,d)}^{(k)}$.
\item M-step \\
With the simulated values, we can now update the estimates for the linear predictor of the incoming intensity, $\hat{\lambda}^{I(k+1)}_{(t,d)}$, as well as the outgoing intensity ${\hat{\lambda}^{R~(k+1)}_{(t,d)}}$.
    \end{enumerate}

\subsection{Bias Correction}\label{sec:biascorrection}
\begin{figure}
    \centering
    \includegraphics[width=\linewidth]{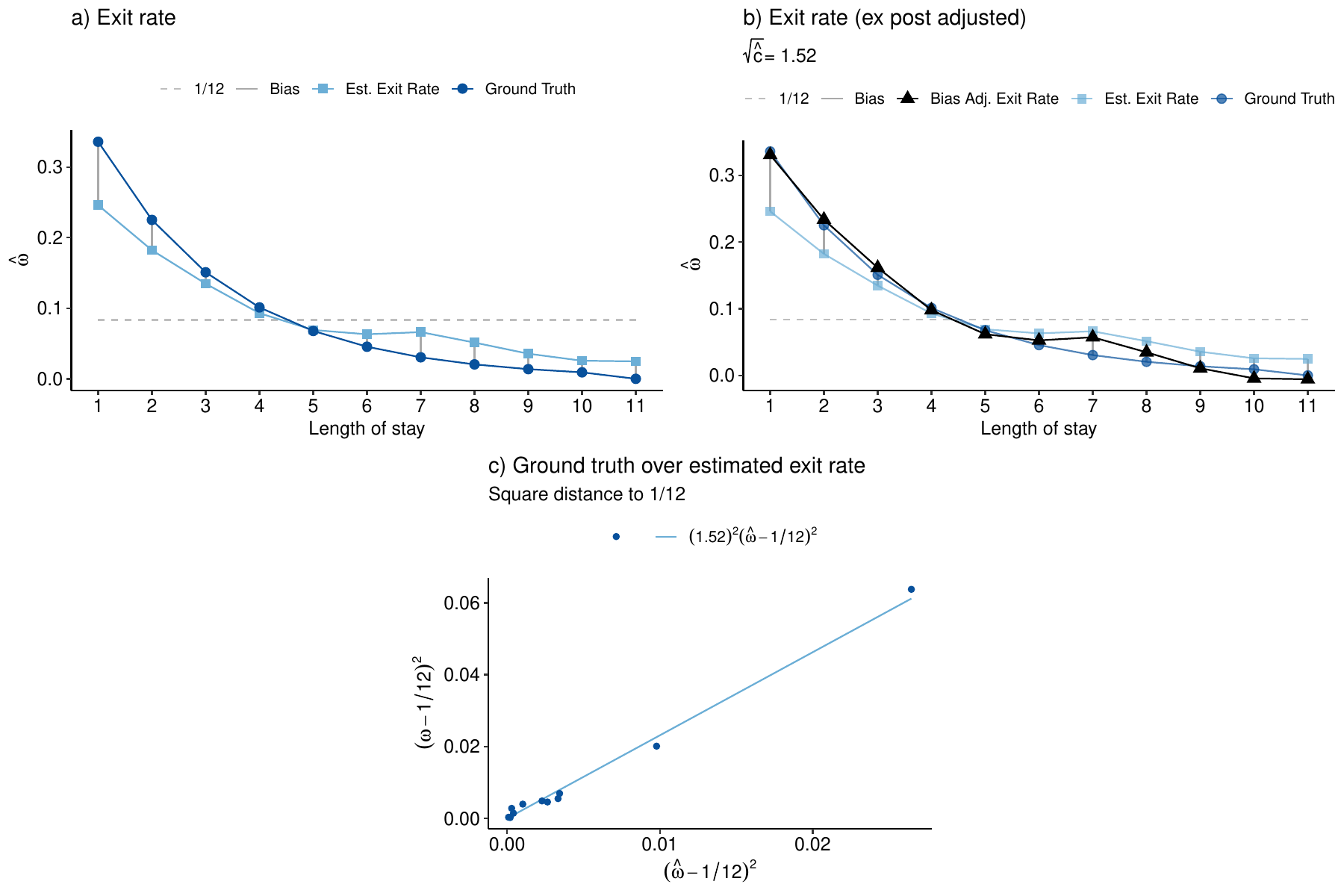}
    \caption{a) Estimated exit rate, $\boldsymbol{\hat{\omega}}$ (light blue squares) and ground truth (dark blue dots) plotted over the length of stay, $l$. The estimated exit rate is not bias adjusted, thus the pull in the estimates from the ground truth towards $1/12$ is shown by the grey vertical lines. (Nota bene: $\hat{\omega}_{12}=1-\sum_{l=1}^{11}\hat{\omega}_{l}$.) b) Illustrated bias adjustment with adjusted estimates $(\boldsymbol{\hat{\hat{\omega}}})$) (black triangles) with an estimated pull $\sqrt{\hat{c}}=1.52$ using the ground truth and the unadjusted exit rate estimate, illustrated in a). c) Square difference between the ground truth exit rate and $1/12$, $(\omega_{l}-1/12)^2$, and square difference between the estimated exit rate and $1/12$, $(\hat{\omega}_{l}-1/12)^2$, with line $y=\hat{c}(\hat{\omega}-1/12)^2$}.
    \label{fig:ExitRateUnAdj}
\end{figure}

By defining the constraints in (\ref{eq:const}) in the estimation of the outgoing intensity (\ref{eq:model-out}), we obtain a prior structure on the coefficient vector $\boldsymbol{\omega}$, which induces a systematic bias. Namely, we find a pull towards a discrete uniform distribution. 
To accentuate this, suppose $Y_{(t,d)}$ is constant over time $Y_{(t,d)}=Y_{(t-1,d)}=Y_{(t-2,d)}=\dots=Y_{(t-n,d)}$, which may occur, for instance, when we encounter an utterly closed system with no incoming nor outgoing units. In this case the vector $\boldsymbol{\omega}$ consists of zero entries, which violates the assumption $\sum_{l=1}^{L} \omega_{l}=1$. The likelihood for $\boldsymbol{\omega}$ is thus flat and the constraints would lead to the estimate $\hat{\omega}_{l} = \frac{1}{L}$, which is evidently biased.
To illustrate the bias problem empirically, we refer to simulated data, which are described in more depth in Section \ref{sec:SimApp}. We apply the sEM outlined in Section \ref{sec:EstApproach}, above. We thus estimate the exit rate without adjustment, for which a pull towards the uniform distribution can be observed. We visualize this in  Figure \ref{fig:ExitRateUnAdj} a), top left-hand side plot. The light blue squares give the final estimates for $\omega_l$. The dark blue dots indicate the ground truth exit rate. The horizontal dashed line is $1/L = 1/12$,  which indicates the probability of a discrete uniform distribution with maximum length of stay equal to $L=12$. We observe an evident pull towards the $1/12$ line.

To correct this bias, we propose bias-corrected estimates of the exit rate. The basic idea relies on the pull towards the $1/L$ line. In Figure 2 c) we plot the squared difference between $\hat{\omega}_l$ as well as $\omega_L$ and $1/L$, respectively. This suggests the  approximate proportionality
\begin{align}
\label{eq:prop}
\left(\hat{\omega}_l - \frac{1}{L}\right)^2 
\approx c \left( \omega_l - \frac{1}{L} \right)^2
\end{align}
for some value $c$. The `best' value of $c$ could be estimated through least squares
\begin{align}
\label{eq:choice-c}
\hat{c}= \mbox{argmin}_c\sum _{l=1}^L \left[\left(\hat{\omega}_{l}-\frac{1}{L}\right)^2-c\left({\omega}_{l}-\frac{1}{L}\right)^2\right]^2.
\end{align}

A bias-corrected version of the estimate is then available by replacing $c$ in (\ref{eq:prop}) by $\hat{c}$ and reversing the pull towards $1/L$. To be precise, we define 
a bias-corrected version through
\begin{align}
    \label{eq:biascorrect}
    \hat{\hat{\omega}}_{l} =
   \left\{ 
   \begin{array}{cl}
    \frac{1}{L}+\sqrt{\hat{c} (\hat{\omega}_{l} -\frac{1}{L})^2} &  \mbox{ for } \hat{\omega}_{l} \geq \frac{1}{L}  \\
    \frac{1}{L}-\sqrt{\hat{c} (\hat\omega_{l} -\frac{1}{L})^2} &  \mbox{for } \hat\omega_{l} \le \frac{1}{L}. 
    \end{array} 
\right.
\end{align}
The resulting bias-corrected estimate is shown in Figure 2 b) on the top right-hand side as black triangles, in addition to the true values and the raw, biased estimates. We see a close concordance with the true values, demonstrating that the bias correction works in the right direction. 

Apparently, looking at formula (\ref{eq:choice-c}), it becomes obvious that the idea is not directly applicable in practice, since we would need the true values $\omega_l$ for $l=1, \ldots ,L$. However, we will utilise the idea and insert an extension to the sEM loop, where we simulate from the $k$-th estimated model and refit the model subesequently. By doing so, we can take the current estimates $\hat{\omega}$ as ground truth and are thereby able to estimate $c$, as described above. The idea is laid out as follows. 

A bias correction is indeed necessary in each iteration step of the sEM algorithm, because a biased estimate of the exit rate $\omega_l$ will induce biased simulations of the incoming patients (sE-step), which in turn will lead to biased estimates of the incoming intensity (M-step). Hence, ignoring the bias creates a chain of problems. To avoid these problems, we propose to extend the sEM-steps 1 and 2 in Section \ref{sec:EstApproach} with a bias correction. 

\begin{enumerate}
\setcounter{enumi}{2}
        \item  
        Simulate data from fitted model\\
        Let $\hat{\boldsymbol{\lambda}}^{I(k+1)}$ and $\hat{\boldsymbol{\lambda}}^{R(k+1)}$ be the estimates resulting after step 1 and 2 in the $k$-th step of the sEM algorithm described in Section \ref{sec:EstApproach}. These estimates are biased and need to be corrected. For the bias correction, the estimates are taken as (current) ground truth. Therefore, 
simulate $\tilde{I}^{(k)}_{(t,d)}$ and $\tilde{R}^{(k)}_{(t,d)}$ using the 
current estimates 
and do \textbf{not} impose $\tilde{I}^{(k)}_{(t,d)}-\tilde{R}^{(k)}_{(t,d)}=\Delta_{(t,d)}$. Instead calculate
    $$
    \tilde{\Delta}^{(k)}_{(t,d)}=\tilde{I}^{(k)}_{(t,d)}-\tilde{R}^{(k)}_{(t,d)}
    $$
    and use these numbers as `simulated observations' from a model, where the parameters are known. 
    
    \item Inner E-Step \textit{(on simulated data)}\\
    Conditional on the `simulated observations', simulate $\check{I}_{(t,d)}$ and $\check{R}_{(t,d)}$ using the current estimates from a Skellam distribution 
    under the condition 
    $$
    \tilde{\Delta}^{(k)}_{(t,d)}\equiv \check{I}_{(t,d)}-\check{R}_{(t,d)}.
    $$
    This can be done as described in Section \ref{sec:EstApproach}. Note,  
$\tilde{\Delta}^{(k)}_{(t,d)}$ here are the simulated differences from step 3 and not the observed data.
    \item Inner M-Step: Outgoing\\
    Use the simulated data from step 4 to obtain estimates $\boldsymbol{\tilde{\omega}}_l$ for $l = 1 , \dots,L$. This can be done as described in Section \ref{sec:EstApproach}.
    \item Bias Correction for Outgoing ($\omega$)\\
    Based on the `raw' estimates $\hat{\omega}^{(k+1)}$ from step 2 and the derived estimates $\tilde{\omega}$ from step 5, calculate the optimal $\hat{c}$ using (\ref{eq:choice-c}), with $\omega_l$ in (\ref{eq:choice-c}) replaced by $\hat{\omega}^{(k+1)}$ and $\hat{\omega}$ replaced by $\tilde{\omega}$.
This yields a bias corrected version for $\hat{\omega}^{(k+1)}$, which is available through (\ref{eq:biascorrect}), that is 
   $$
   \hat{\hat{\omega}}^{(k+1)}_{l}=\begin{cases}
 \frac{1}{L}+\sqrt{\hat{c}(\hat{\omega}^{(k+1)}_{l}-\frac{1}{L})^2}, ~~~~ \hat{\omega}^{(k+1)}_l \geq \frac{1}{L}\\
 \frac{1}{L}-\sqrt{\hat{c}(\hat{\omega}^{(k+1)}_{l}-\frac{1}{L})^2}, ~~~~ \hat{\omega}^{(k+1)}_l< \frac{1}{L}
\end{cases}
$$
\item Bias Correction for Incoming ($\lambda^{I}$) \\
Simulate incoming and outgoing patients again, like in step 1, but now using
the current (raw) estimates $\hat{\boldsymbol{\lambda}}^{I~(k+1)}$ and the bias-corrected estimates $\hat{\hat{\omega}}^{(k+1)}$ and conditional on the observed data
   $$
    \Delta_{(t,d)}\equiv \tilde{I}^{(k)}_{(t,d)}-\tilde{R}^{(k)}_{(t,d)},
    $$
Note, this is like the original step 1 in the sEM algorithm, but a bias-corrected version replaces the exit rates. \\ Use the simulated incoming patients to obtain a bias-corrected version $\hat{\hat{\boldsymbol{\lambda}}}^{I~(k+1)}$.
\item Concluding the loop \\
Replace $\boldsymbol{\hat{\omega}}^{(k+1)}$ by 
$\boldsymbol{\hat{\hat{\omega}}}^{(k+1)}$ and $\boldsymbol{\hat{\lambda}}^{I~(k+1)}$ by $\boldsymbol{\hat{\hat{\lambda}}}^{I~(k+1)}$ and proceed with step 1 in the sEM algorithm. 
\end{enumerate}
In the application, we suggest extending steps 1 and 2 of the sEM loop with the extra steps 3 to 8 not immediately, but only after some `burn-in' phase. This accelerates the estimation process. 

\subsection{Inference}
Given the application of the sEM we can use the variability of the estimates within the sEM chain to adjust for the underestimated variance, as given by \citep{rubin1976inference}. Let therefore $\boldsymbol{\beta}$ denote the parameter vector with all model parameters stacked together. 
We use the variance estimation 
\begin{equation}\label{eq:VarianceEstimate}
\hat{\Sigma}_{\boldsymbol{\beta}}=\frac{\sum_{k=k'}^K\Sigma_{\boldsymbol{\beta}}^{(k)}}{K-k'}+\frac{\sum_{k=k'}^K(\boldsymbol{\beta}^{(k)}-\boldsymbol{\bar{\beta}})(\boldsymbol{\beta}^{(k)}-\boldsymbol{\bar{\beta}})^T}{K-k'-1},
\end{equation}
with $\Sigma^{(k)}$ being the covariance matrix estimated at the $k^{th}$ iteration, 
$\boldsymbol{\bar{\beta}}$ being the mean (or median in case of outliers) estimate of the last $K-k'$ runs of the column coefficient vector $\boldsymbol{{\beta}}$, with $k'$ being a starting point at which convergence is assumed to have occurred. 
The estimated covariance matrix for the model on incoming units, (\ref{eq:incom_est}), is a standard estimation. For the model on outgoing units, (\ref{eq:outg_est}), we take the inverse of (\ref{eq:FishInf}) as an estimate for the covariance matrix. For simplicity, we assume the incoming and outgoing units to be independent. 

\section{Simulation}\label{sec:SimApp}

We simulate a data example, which is aimed to emulate the real data closely. We simulate $200$ districts, $d$, for which we observe data at $200$ time points, $t$. This results in $40.000$ observations. 
We then simulate two covariates from which the incoming units are simulated, as seen in (\ref{eq:IncSim}). 
\begin{align}\label{eq:IncSim}
\nonumber x1_{(d)}&\sim \mbox{Gamma}(0.1,0.5),~~ \mbox{(Nota bene: varying over districts, constant over time)}\\
x2_{(t,d)}&\sim \mbox{Gamma}(1,3),\\
\nonumber \lambda^I_{(t,d)}&=\exp(0.5+\mbox{x1}_{(d)}+0.2 \mbox{ x2}_{(t,d)})\\
\nonumber I_{(t,d)}&\sim \mbox{Poisson}(\lambda^I_{(t,d)})\\
\nonumber ~~ \forall~~ &t \in \{1,\dots, 200\},~~ d \in \{1,\dots, 200\}.
\end{align}

For the simulation setup, we choose the maximum length of stay to be $L = 10$. The probability mass function is 
\begin{align}
    \pi_{l}=P(L=l)&=\frac{\exp(-0.4 l)}{\sum_{s=1}^{10} \exp(-0.4 s)}.
\end{align}
From this, we now simulate the outgoing number of units in a slightly different way to the estimation procedure. 
Namely, let $(\pi_{1}, \dots, \pi_{10})$ and for each incoming patient $i_{(t,d)} \in \{ 1, \ldots , I_{(t,d)}\}$ at time $t$ and district $d$ we simulate a length of stay, $l_{i_{(t,d)}}$, from $(\pi_{1}, \dots, \pi_{10})$. Then 
\begin{equation}\label{eq:OutgSim} 
R_{(t,d)}=\sum_{l=1}^L\sum_{i=1}^{I_{(t-l,d)}} {1}(l_{i_{(t,d)}}=l),
\end{equation}
with ${1}(.)$ denoting the indicator function.
To summarise, the number of outgoing units, at time point $t$ and district $d$, is the sum of units which have previously come in $l$ days before.  

From (\ref{eq:IncSim}) and (\ref{eq:OutgSim}) we obtain the difference,
\begin{equation} \label{eg:SimDiff}
\Delta_{(t,d)}=I_{(t,d)}-R_{(t,d)}.
\end{equation}

Once the data are generated, the sEM is applied for $400$ iterations. For different starting values, the sEM would take a different number of iterations until convergence is observed. However, we conjecturally observe a convergence rather quickly, maximally after $100$ iterations. In Appendix \ref{app:Converg}, we observe that the likelihood has reached some convergence after around $50$ iterations of the sEM. We summarise the results for the last $200$ runs of the applied sEM, by the median of respective point estimates and the estimated standard deviation. The M-Step comprises the estimation of the incoming intensity parameter and the exit rates. For the exit rate, we select a maximum lag $L$ that exceeds the true lag used in the simulation. This should mirror a plausible estimation strategy, where, for estimation, one sets the maximum lag large enough, potentially larger than needed. To be specific, we set the maximum lag for fitting to be $12$. The estimate of the incoming intensity parameter is given by

\begin{equation}\label{eq:IncEstModel}
{\hat{\lambda}_{(t,d)}^I}=\exp(\hat{\beta}_0+\hat{\beta}_1\mbox{x1}_{(d)}+\hat{\beta}_2\mbox{x2}_{(t,d)}).
\end{equation}

\begin{figure}
    \centering
    \includegraphics[width=0.7\linewidth]{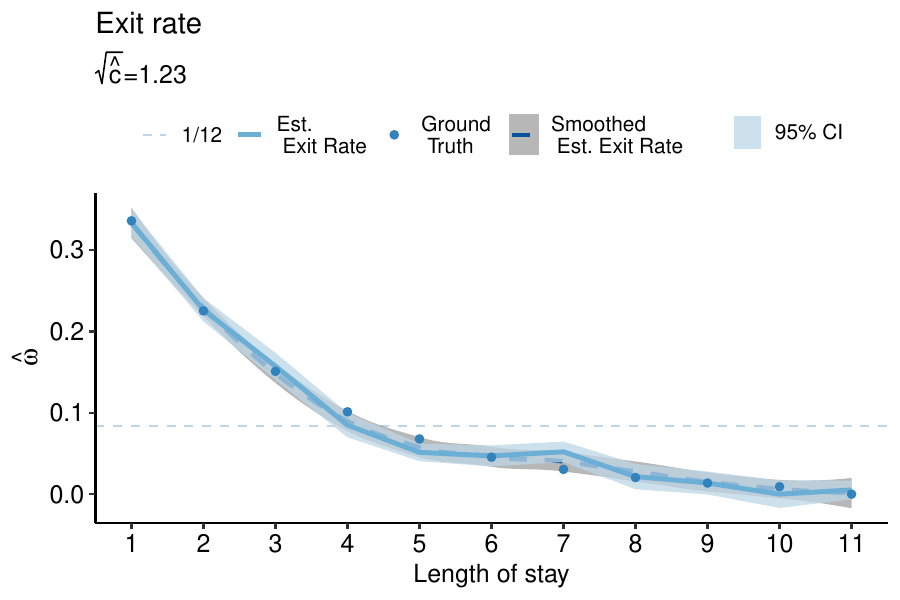}
    \caption{Estimated exit rate over the length of stay (denoted lag) with $95\%$ confidence interval (of the last $200$ runs of the sEM) against ground truth, with $\hat{\omega}_{12}=1-\sum_{l=1}^{11}\hat{\omega}_{l}$.}
    \label{fig:ExitRateSimPois}
\end{figure}

\begin{figure}
    \centering
    \includegraphics[width=0.8\linewidth]{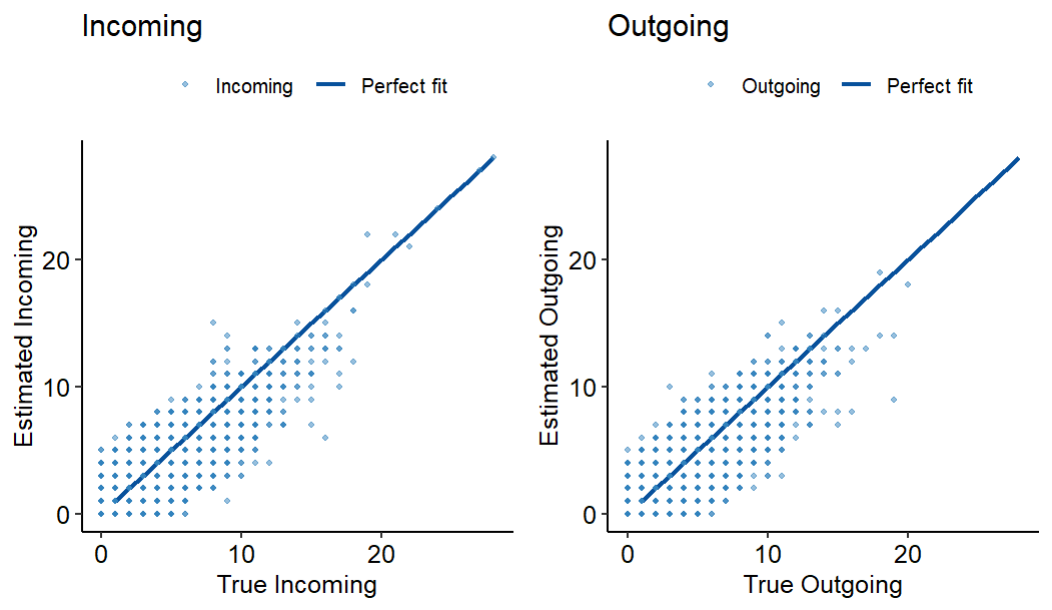}
    \caption{Estimated incoming and outgoing number of units.}
    \label{fig:IncomOutSim}
\end{figure}

In Figure \ref{fig:ExitRateSimPois} and Table \ref{tab:ResultsCoefSim} the median of the point estimates and the standard deviation, the square root of the variance estimate as given in (\ref{eq:VarianceEstimate}), are displayed, where the median of the simulated incoming and outgoing units are displayed in Figure \ref{fig:IncomOutSim}.

Table \ref{tab:ResultsCoefSim} shows the estimated and true effects of the covariates on the incoming units. We observe that the estimates approximate the ground truth for both coefficient estimates. However, we observe a somewhat larger deviation for the estimated intercept. We will get back to this point in a second simulation setup below. The true and estimated exit rates, shown in Figure \ref{fig:ExitRateSimPois}, evidence an estimation close to the ground truth for all estimates of the exit rate, with some slight deviation from the $95\%$ confidence interval at lag $5$ and lag $7$. Note this is just one simulation and overinterpretation should be avoided. Therefore, we additionally fit a smooth fit to estimated exit rates, which mitigates the random deviations from the true exit rates.  

\begin{table}[h]
    \centering
    \begin{tabular}{lccc}
        \hline
        Parameter & Estimate & Std. Dev. & Ground Truth \\
        \hline
        ${\beta}_0$ & 0.3788 & 0.0089 & 0.5 \\
        ${\beta}_1$ & 0.9828 & 0.0167 & 1 \\
        ${\beta}_2$ & 0.2149 & 0.0036 & 0.2 \\
        \hline
    \end{tabular}
    \caption{Results of coefficients against ground truth.}
    \label{tab:ResultsCoefSim}
\end{table}

Though this simulation shows that the model is able to estimate the true underlying incoming and outgoing units, as well as the true coefficients, in practice, there is likely to be overdispersion in inflow and outflow relative to the Poisson model assumption. We therefore extend the simulation process by generating inflow data from a Negative Binomial distribution (\ref{eq:IncSimNegBin}), instead of a Poisson distribution. In doing so, we retain the true incoming coefficients for the expected value of incoming units, as specified in (\ref{eq:IncSim}):

\begin{align}\label{eq:IncSimNegBin}
I_{(t,d)}&\sim \mbox{Negative-Binomial}(\lambda^I_{(t,d)}, \theta).
\end{align}

In (\ref{eq:IncSimNegBin}) the variance assumption extends from that of the Poisson distribution to
\begin{equation}\label{eq:VarianceNB}
\mathbb{V}_{NB}(I_{(t,d)})=\lambda^I_{(t,d)}+\frac{\left(\lambda^I_{(t,d)}\right)^2}{\theta}.
\end{equation}

We simulate data for different values of $\theta$, with ${\theta}_1=0.5,{\theta}_2=1,{\theta}_3=5,~{\theta}_4=10$.  The simulated overdispersion decreases with increasing $\theta$. For each $\theta$, we again simulate $200$ districts and $200$ time points, resulting in $40{,}000$ observations per data set, and estimate inflow and outflow analogously to the previous setup.

\begin{table}[h]
\centering
\caption{Estimated coefficients from the $200^{th}$ to the $400^{th}$ of misspecified models compared to the true values.}
\label{tab:TableCoefficients}
\begin{tabular}{l|rrr}
 & $\beta_0$ & $\beta_M$ & $\beta_N$ \\
\hline
True         & 0.500 & 1.000 & 0.200 \\
\hline
$\theta=0.5$ & 1.269 & 1.925 & 0.099 \\
$\theta=1$   & 0.937 & 1.662 & 0.137 \\
$\theta=5$   & 0.499 & 1.406 & 0.178 \\
$\theta=10$  & 0.487 & 1.025 & 0.219 \\
\hline
Poisson  & 0.379 & 0.983 & 0.215 \\
\hline
\end{tabular}
\label{tab:ResultsCoefSimALL}
\end{table}

Table \ref{tab:ResultsCoefSimALL} reports the estimated and true covariate effects for each simulated data set, including the Poisson-based simulation for comparison. We observe that the estimates approach the ground truth as overdispersion decreases.
We also refer to Appendix \ref{app:PredInOut}, where we show simulated incomings, from one of the last stochastic E-steps, plotted against the true incomings, based on the simulations. Overall, the models’ estimates tend to approximate the true coefficients more closely as overdispersion diminishes.

\begin{figure}
    \centering
    \includegraphics[width=0.7\linewidth]{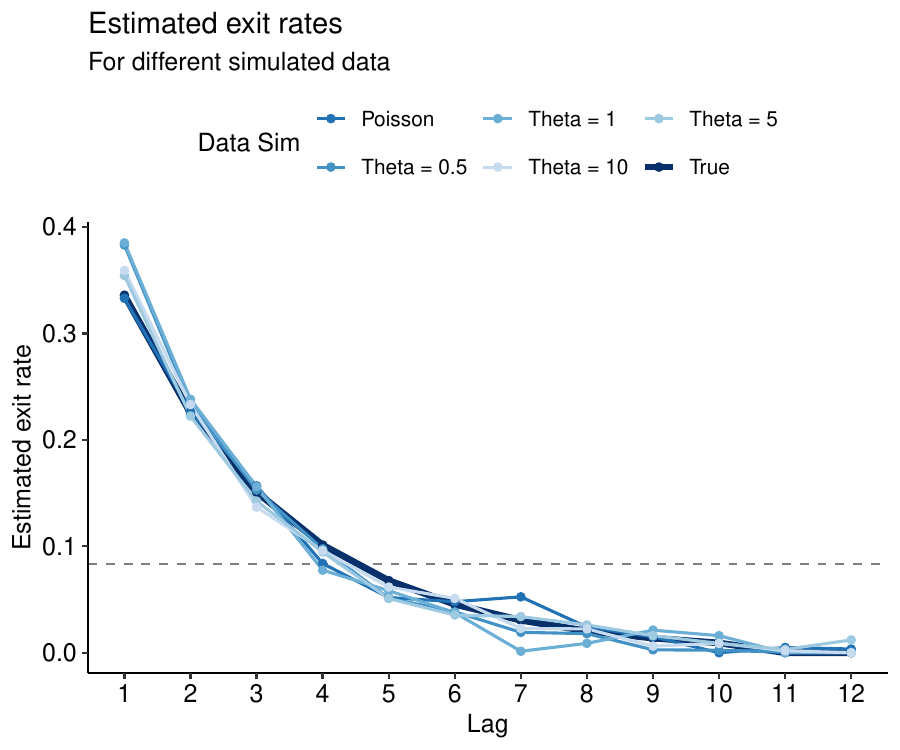}
    \caption{Estimated exit rate over the length of stay (denoted lag) for models applied to data simulated, with incoming units simulated from a Poisson distribution and Negative-Binomial distributions, with ${\theta}_1=0.5,~{\theta}_2=1,~{\theta}_3=5,~{\theta}_4=10$.}
    \label{fig:ExitRateSimALL}
\end{figure}

Looking at the exit rate, which is the primary focus of interest, we see from Figure \ref{fig:ExitRateSimALL} that overdispersion does not disturb roughly consistent estimation.  The estimates of the exit rate all show a similar performance.

\section{Results}\label{sec:CovApp}
With the above prerequisites, we are now able to apply our model to the ICU data.
For stability in our estimation, we first apply the sEM, as a `pre-run', to the data for a total $200$ iterations, without conducting any bias adjustment. Said `pre-run' renders results which are assumed to be in a reasonable range for starting values of the sEM with bias adjustment, i.e. actually used in estimation results. The sEM with bias adjustment runs for another $150$ iterations. The final results are summarized over the last $100$ runs. The log-likelihood over the initial $200$- unadjusted- iterations and the subsequent $150$- adjusted- iterations are shown in Appendix \ref{app:Converg}. For reference, the linear predictor for the incoming intensity is given in (\ref{eq:IncEstCov}).

In Table \ref{tab:coef_fix} we see the results of the estimated effects of the infection rates and the weekday effects. We see that the estimated effect of the lagged infection rates of the `35-59' year olds is largest, which agrees with the findings of our first paper, see \citep{rave2023skellam}. The estimated weekday effects further agree with our initial findings, where we estimate to see less incoming patients into the ICU on weekends, compared to Fridays, and more during weekdays, again, compared to Fridays. Contextually, one might argue that the severity of a disease might not care about the day of the week. However, this might be explained by internal movements within a treatment centre, where severe cases might first be treated in an Emergency Room (ER), and only be moved to the ICU, once the appropriate personnel has authorised it.  

\begin{figure}
    \centering
    \includegraphics[width=\linewidth]{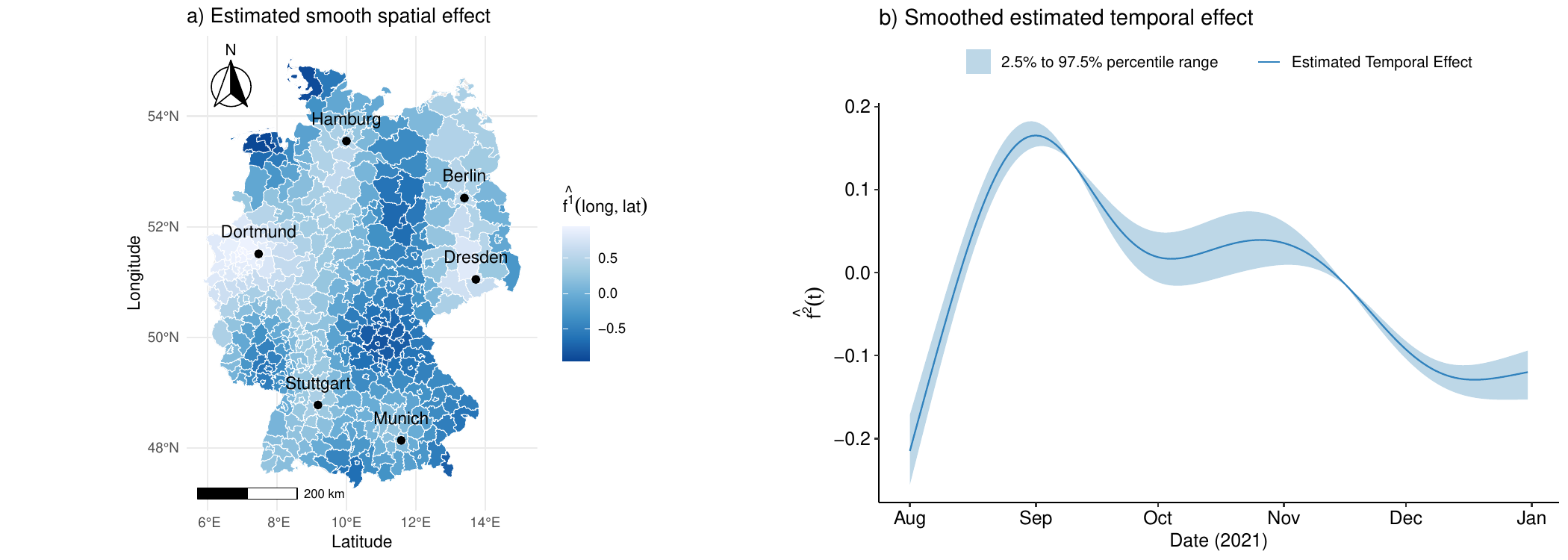}
    \caption{a) Estimated smooth function over space b) Estimated smooth function over time.}
    \label{fig:SpatialEffCov}
\end{figure}

In Figure \ref{fig:SpatialEffCov} a), we see the estimated spatial effects, were we observe an increase, to varying degrees, in and around large cities, such as Dortmund, Hamburg, Dresden, Berlin, Stuttgart and Munich. This also agrees with the findings of our earlier work. Contextually, in the centralised health care system of Germany, we tend to have more ICU capabilities in the cities, which leads to ICU patients from surrounding rural areas typically being treated in near cities, rather than in their district. Particularly, during the COVID-19 pandemic, city hospitals were usually the treatment centres with treatment capabilities for isolation and respiration of COVID-19 patients. So rather than directly inferring that the severity of the disease being stronger in urban environments, the factor of the hospitalisation logistics may also be a driving factor here.

In Figure \ref{fig:SpatialEffCov} b), we observe the estimated smooth function over time. It is wrapped by the $2.5^{th}$ and $97.5^{th}$ percentile of the estimated smooth functions of the $100$ included sEM iterations. We observe an initial increase in the estimated smooth function until September, 2021, with a subsequent sharp decrease, a slight pick up from October until November and a following decrease, which seems to pick up again in the end of December, 2021. The interpretation of the estimated temporal effect is, as all other interpretations, conducted ceteris paribus. Thus, we estimate an increasing admittance to the ICU until September, which cannot be entirely explained by the other covariates included in our estimation. This is followed by a subsequent fall in ICU admittance, likewise not explained by the other estimated effects.  

\begin{table}[h]
    \centering
    \begin{tabular}{lcc}
        \hline
        \textbf{Covariates} & \textbf{Estimates} & \textbf{Std. Dev.} \\
        \hline
        Intercept & -2.811 & 0.040 \\
        Infection Rate 35-59 & 0.545 & 0.023 \\
        Infection Rate 60-79 & 0.098 & 0.024 \\
        Infection Rate 80+ & 0.112 & 0.011 \\
        Monday & 0.105 & 0.022 \\
        Tuesday & 0.045 & 0.023 \\
        Wednesday & 0.033 & 0.023 \\
        Thursday & 0.071 & 0.022 \\
        Saturday & -0.018 & 0.023 \\
        Sunday & -0.086 & 0.023 \\
        \hline
    \end{tabular}
    \caption{Estimated coefficients on inflow of ICU patients.}
    \label{tab:coef_fix}
\end{table}

In Figure \ref{fig:RKIvsEst}, we show the estimated incoming patients aggregated to Bundesland level, plotted against the ``Erstaufnahmen'' (incoming) patients, reported by the  \citep{DIVI_REPO}. We see that our model underestimates the number of incoming patients in Berlin, which would fit intuition, following our centralised health care system interpretation of the estimated spatial effects. For a better visual impression, we included a smooth estimate of the exit rates. 

\begin{figure}
    \centering
    \includegraphics[width=0.7\linewidth]{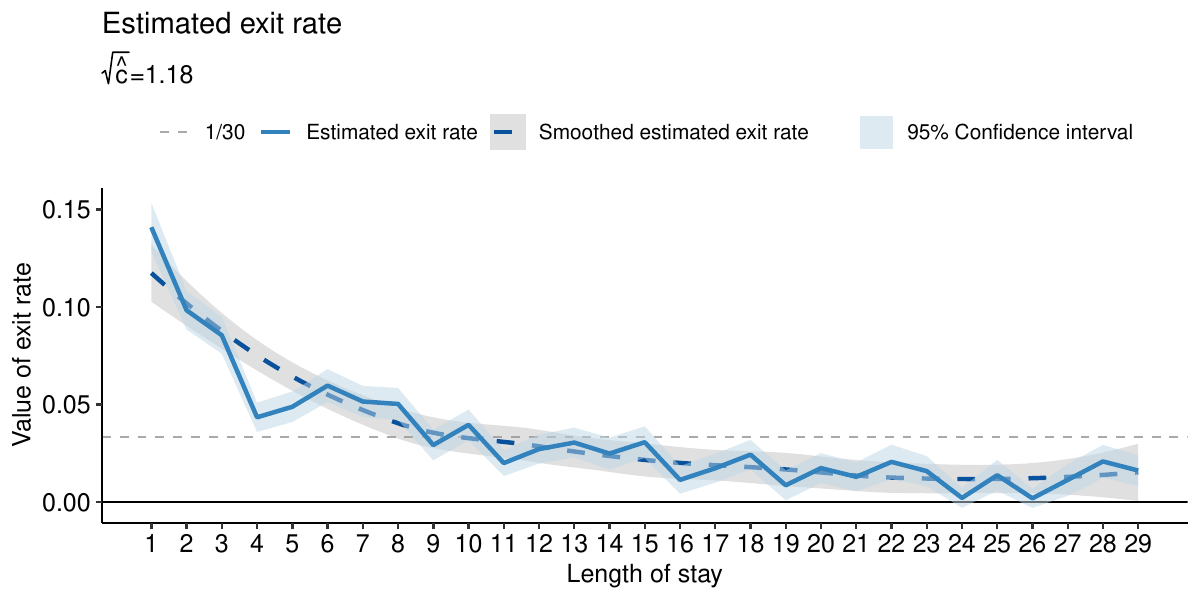}
    \caption{Estimated exit rate with $95\%$ confidence interval (of the last 100 runs of the sEM) with smoothed estimate over exit rates.}
    \label{fig:ExitRateCov}
\end{figure}

Figure \ref{fig:ExitRateCov} shows the estimated exit rates up until a maximum of a 30 day lag. We estimate a sharp decline in the estimated exit rate along the initial 16 days, and a subsequent slough off thereafter. More specifically, we see an estimate of around $13\%$ of ICU patients with COVID-19 leaving after one day, $50\%$ of patients are estimated to have left by their $6^{th}$ day in the ICU and $80\%$ of COVID-19 patients are estimated to have left the ICU by the 17$^{th}$ day. Finally, $90\%$ of COVID-19 patients are estimated to have left after 22 days.

Inspecting the \citep{DIVI_REPO} repository, one may discover, that since 2021 data on the number of admitted ICU patients with COVID-19 have been published. However, the most granular these data are published, are on state level (there are 16 states in Germany), while our data are on the district level, which make up each of the respective counties to which they belong. We may therefore aggregate our estimated admitted ICU patients and compare them with the data reported by the RKI. In Figure \ref{fig:RKIvsEst}, we plot our aggregated estimation against the RKI reported data. Specifically in Berlin and Brandenburg (titles marked by the blue outline), we observe a clear deviance. This may be due to hospital logistics, which we have not included in our analysis. The health care system in Germany induces that treatment facilities in cities tend to be more equipped to treat patients in need of specialised care, such as isolation for patients infected with COVID-19. A short outline of this principle during the COVID-19 pandemic and the planned cooperation between counties is given by \citep{Aertze2020}. We thus underestimate the number of admitted ICU patients with COVID-19 in Berlin, as we suspect that many of which will have been moved from surrounding counties, such as Brandenburg, where we overestimate the number of admitted ICU patients.   

\begin{figure}
    \centering
    \includegraphics[width=\linewidth]{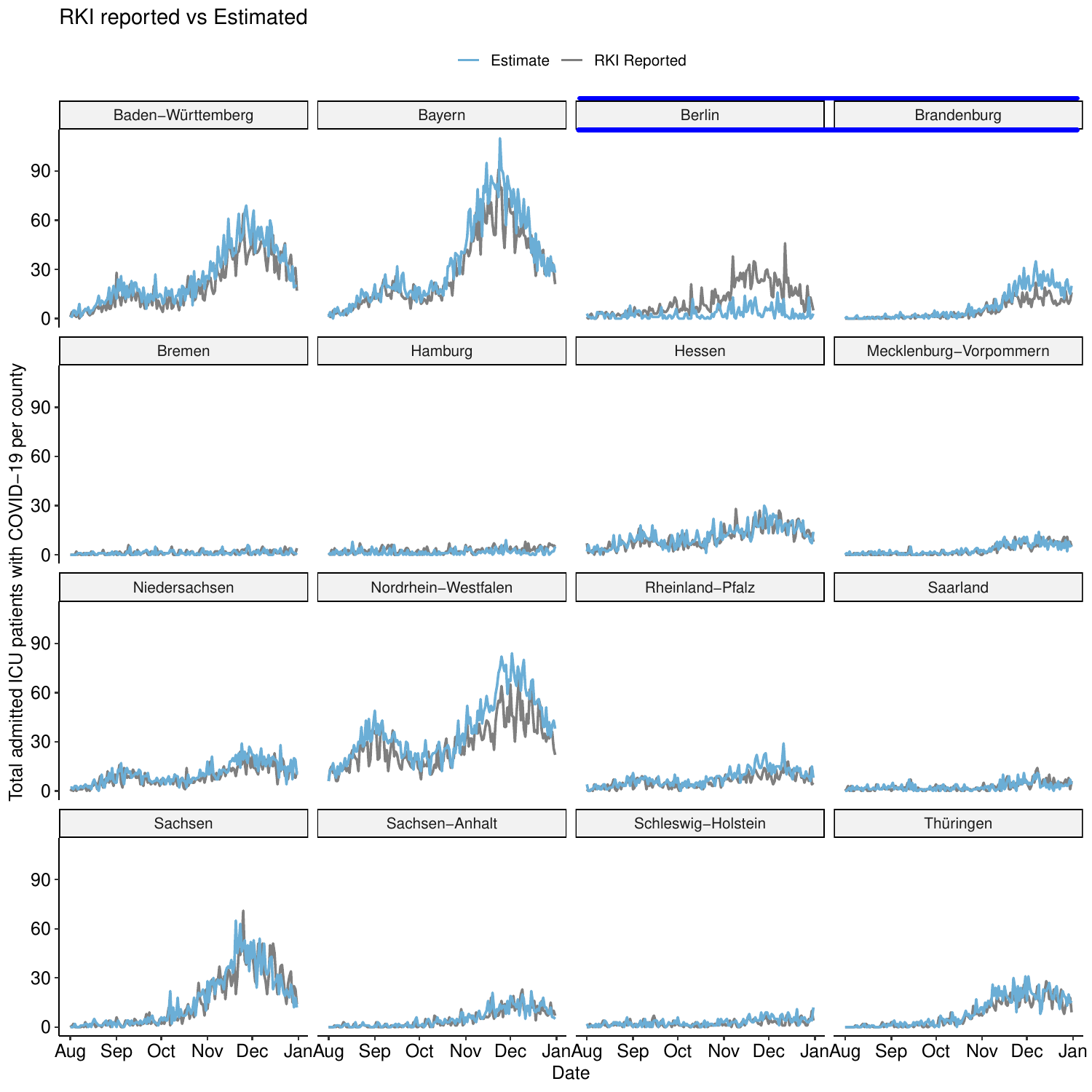}
    \caption{RKI reported admitted ICU patients with COVID-19 over time plotted against the estimated admitted ICU patients.}
    \label{fig:RKIvsEst}
\end{figure}

\section{Discussion}\label{sec:Diss}
Our approach demonstrates that we can extract information on underlying inflow and outflow processes by observing current snapshots of a system only. We also show how to include further covariates which influence the incoming intensity. As remarked in the introduction, the idea can be extended to similar data constellations. For example, the field of population dynamics would benefit from our approach in that herd inflow and outflow are often expensive to record continuously over a long period of time. Our model is able to circumvent this predicament elegantly by including information on the inflow.

In the estimation of the length of stay, we draw on a `non-standard' estimation process through the bias adjustment. There is a possibility that the bias is merely mitigated, but still present, thus implying we underestimate the variance. In further work, one could refine the approach to adjust for bias in the variance estimation and thereby achieve better coverage of the estimates. 

Despite the advantages of our approach, we do encounter some challenges when fitting the sEM. We have a clear disadvantage in the running time of the algorithm. This is likely optimisable in our particular model, however, only to a certain degree, with a clear limitation being the stochastic nature of the algorithm.  All in all, the estimation requires iterative simulations due to the sequential pattern of the model. This leads inevitably to heavy computation. 

A further possible extension to the model arises from the context of the COVID-19 ICU data. We do not differentiate between patients who were moved to Intermediate Care Units (IMCU) or other units within the hospital and patients who die during their stay at the ICU. We also do not take the movement of patients between counties into account. It is therefore likely that our model predicts the number of admitted ICU patients by district of origin well, but does not take patients' placement between districts into account and therefore deviates from the RKI-reported data.

Our approach allows us to obtain information about data which were not made public at the time of the analysis. Apparently, for practical purposes, it is certainly better to record the original data and omit the modelling exercise pursued in this paper. Meaning that in our view it seems advisable to enable any data system to incorporate the true data on incoming and outgoing patients in ICU units. 

\section*{Declaration of Conflicting Interests}
The authors declared no potential conflicts of interest with respect to the research, authorship, and/or publication of this article.

\section*{Data Availability Statement}
The data that support the findings of this study are openly available at
\\
\href{https://robert-koch-institut.github.io/Intensivkapazitaeten_und_COVID-19-Intensivbettenbelegung_in_Deutschland/}{\url{https://robert-koch-institut.github.io/Intensivkapazitaeten\_und\_COVID-19-Intensivbettenbelegung\_in\_Deutschland/}} 
\\
 and \href{https://robert-koch-institut.github.io/COVID-19_7-Tage-Inzidenz_in_Deutschland/}{\url{https://robert-koch-institut.github.io/COVID-19\_7-Tage-Inzidenz\_in\_Deutschland/}}.

\newpage

\bibliographystyle{plainnat}
\bibliography{Bibliography}

\newpage

\appendix

\section{Score function and Fisher Information} \label{app:scorefisher}
We derive the approximate score function from (\ref{eq:partlike}),    
    \begin{equation}
    s(\hat{\omega}_l^{(k)})=
    \frac{\partial l^R _P(\boldsymbol{\omega})}{\partial {\omega}_l}\bigg|_{\boldsymbol{\omega}=\hat{\boldsymbol{\omega}}^{(k)}}=\sum _{t=1}^T\sum _{d=1}^D \Big(({I}_{(t-l,d)}-{I}_{(t-L,d)})\Big(\frac{R_{(t,d)}}{\sum_{l=1}^{L}\hat{\omega}_{l}^{(k)}{I}_{(t-l,d)}}-1\Big)\Big) 
    \end{equation}
    and the second-order derivative
    \begin{equation}
    \label{eq:FishInf}
    \mathcal{I}_{jk}(\hat{\boldsymbol{\omega}}^{(k)})= \frac{\partial l^R_P(\boldsymbol{\omega})^2}{\partial \omega_j\partial \omega_k}\bigg|_{\boldsymbol{\omega}=\hat{\boldsymbol{\omega}}^{(k)}}=-\sum _{t=1}^T\sum _{d=1}^D R_{(t,d)}\frac{({I}_{(t-l,d)}-{I}_{(t-L,d)})({I}_{(t-k,d)}-{I}_{(t-L,d)})}{(\sum_{l=1}^L\hat{\omega}_{l}^{(k)}{I}_{(t-l,d)})^2},
    \end{equation}
for $l=\{1,\dots, L-1\}$, $j=\{1,\dots, L-1\}$ and $k=\{1,\dots, L-1\}$. 
These terms are derived to determine the second-order approximation (\ref{eq:partiallike}).

\section{Truncated joint probability mass function} \label{app:jPMF}
First, we define a reasonable range $[0, I_{max}]$ of probable income values $I_{(t,d)}$, such that $ \mbox{p}(I_{(t,d)}\geq I_{max},  ~R_{(t,d)}\geq I_{max}-\Delta_{(t,d)}| \lambda^I_{(t,d)}, \lambda^R_{(t,d)})\approx 0$ . Then we calculate the conditional probability 
   \begin{align}
   \label{eq:truncPMF_appendix}
    p(I_{(t,d)}=i,  ~&R_{(t,d)}=i-\Delta_{(t,d)}| I_{(t,d)} \leq I_{max}; \lambda^I_{(t,d)}, \lambda^R_{(t,d)})\\
    \nonumber
    &= \lim_{Q\rightarrow\infty} \frac{\exp(-\lambda^I_{(t,d)}){[\lambda^I_{(t,d)}]}^{i}~ \exp(-\lambda^R_{(t,d)})[\lambda^R_{(t,d)}]^{i-\Delta_{(t,d)}} (i! (i-\Delta_{(t,d)})!)^{-1}}{\sum_{j= 0}^{Q} [\exp(-\lambda^I_{(t,d)}){[\lambda^I_{(t,d)}]}^j\exp(-\lambda^R_{(t,d)}){[\lambda^R_{(t,d)}]}^{j-\Delta_{(t,d)}} (j! (j-\Delta_{(t,d)})!)^{-1}]}\\
    \nonumber
    &\approx \frac{\exp(-\lambda^I_{(t,d)}){[\lambda^I_{(t,d)}]}^{i}~ \exp(-\lambda^R_{(t,d)})[\lambda^R_{(t,d)}]^{i-\Delta_{(t,d)}} (i! (i-\Delta_{(t,d)})!)^{-1}}{\sum_{j= 0}^{I_{max}} [\exp(-\lambda^I_{(t,d)}){[\lambda^I_{(t,d)}]}^j\exp(-\lambda^R_{(t,d)}){[\lambda^R_{(t,d)}]}^{j-\Delta_{(t,d)}} (j! (j-\Delta_{(t,d)})!)^{-1}]},
        \end{align} 

    $\forall~i~\in \{0, \dots, I_{max}\}$. For conciseness, we omitted the indicator for sampling at the $k$-th iteration.

\textit{Nota bene: The bias correction need not be conducted at every iteration of the sEM. Particularly, the estimation of parameters where the likelihood is multimodal, or the assumed model is highly complex. A suggested solution is to conduct a sEM, without the bias correction until convergence is reached, and then use the obtained estimates as starting values for conducting an sEM with bias correction.}

\section{Convergence} \label{app:Converg}

\begin{figure}[H]
    \centering
    \includegraphics[width=0.7\linewidth]{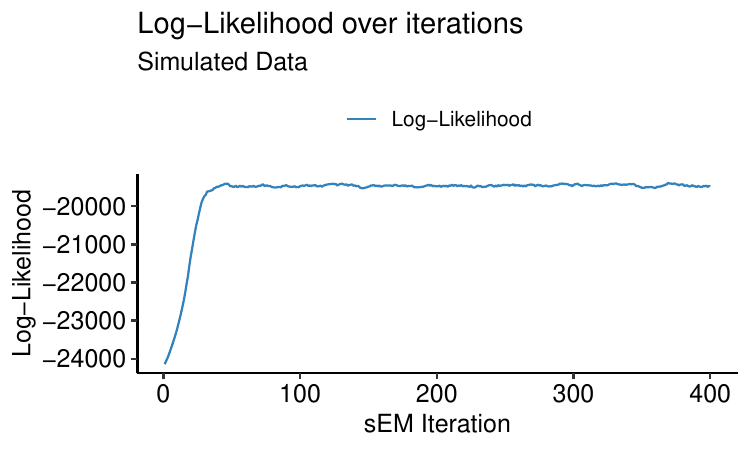}
    \caption{Log-Likelihood over $400$ iterations of the sEM applied to simulated data.}
    \label{fig:LogLikeSim}
\end{figure}

\begin{figure}[H]
    \centering
    \includegraphics[width=\linewidth]{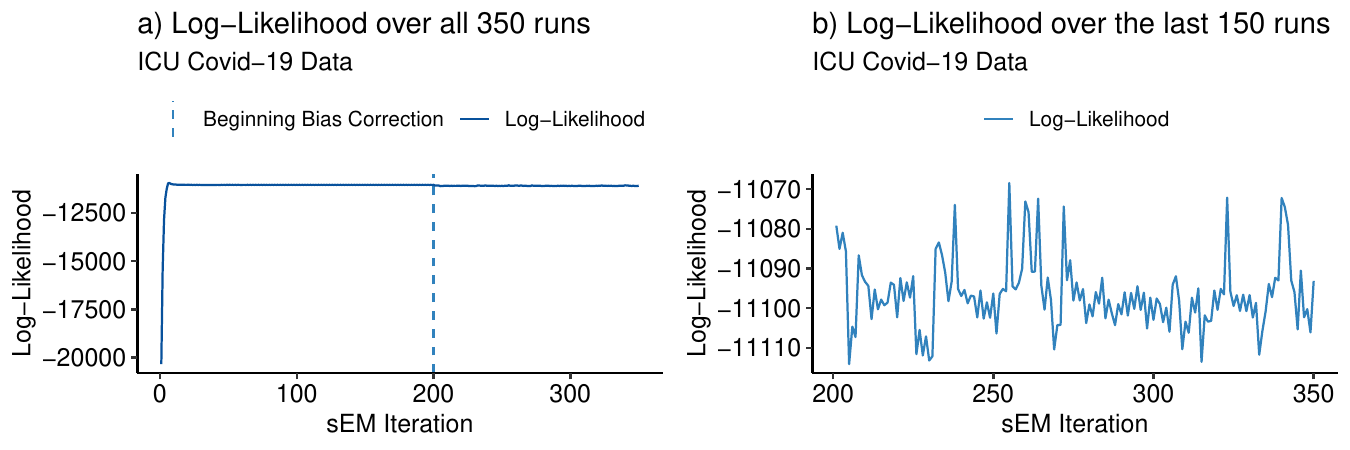}
    \caption{a) Log-Likelihood over $350$ iterations of the sEM applied to ICU COVID-19 data (initial $200$ iterations `burn-in' without bias correction, subsequent $150$ iterations are implemented using bias correction). b) Log-Likelihood zoomed in over the last $150$ iterations of the sEM applied to ICU COVID-19 data.\\
    NB: The y-axes in a) and b) are on different scales.}
    \label{fig:LogLikeCov}
\end{figure}

\section{Simulation-Predicted Incoming and Outgoing} \label{app:PredInOut}

\begin{figure}
    \centering
    \includegraphics[width=0.5\linewidth]{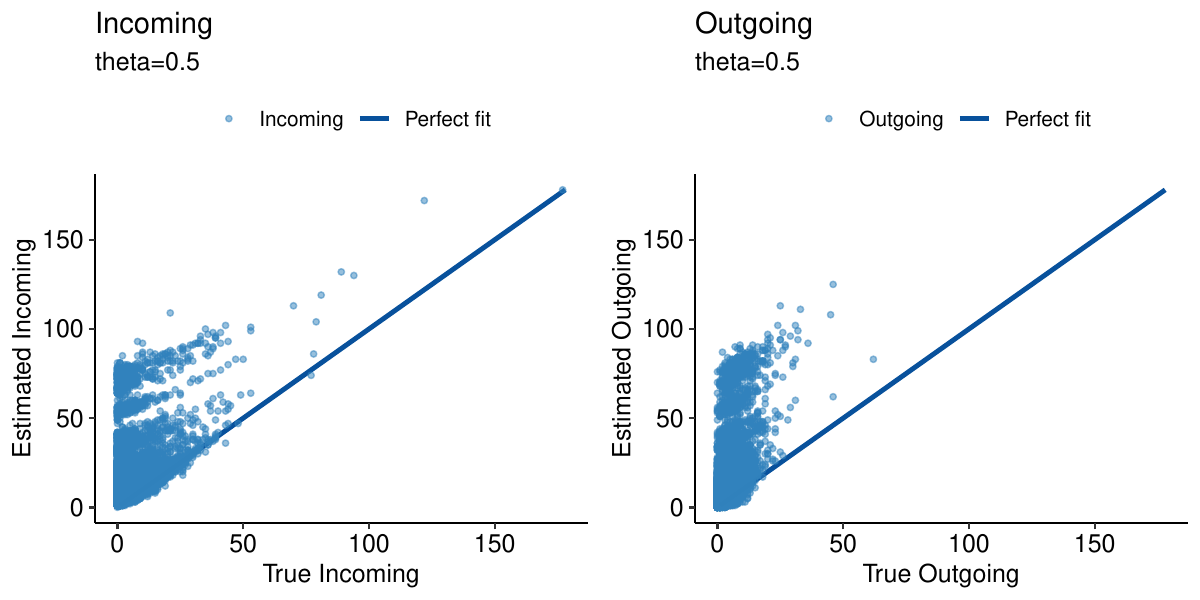}
    \caption{The estimated inflow and outflow for data with chosen data with Negative Binomial- $\theta=0.5$.}
    \label{fig:IncOut_Mis_05}
\end{figure}

\begin{figure}
    \centering
    \includegraphics[width=0.5\linewidth]{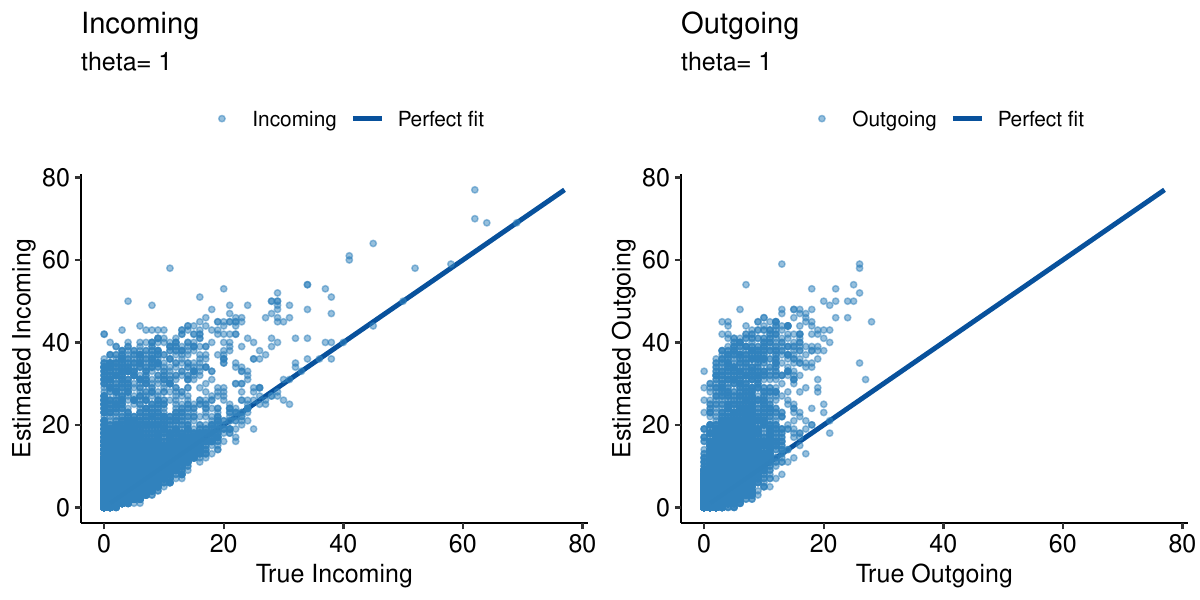}
    \caption{The estimated inflow and outflow for data with chosen data with Negative Binomial- $\theta=1$.}
    \label{fig:IncOut_Mis_1}
\end{figure}

\begin{figure}
    \centering
    \includegraphics[width=0.5\linewidth]{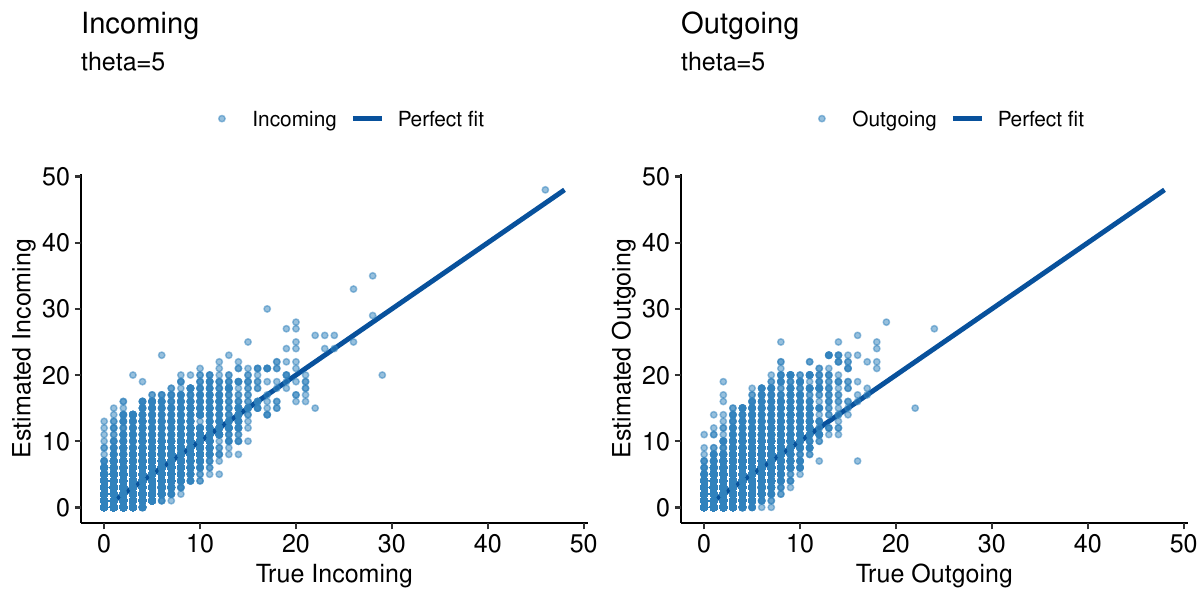}
    \caption{The estimated inflow and outflow for data with chosen data with Negative Binomial- $\theta=5$.}
    \label{fig:IncOut_Mis_5}
\end{figure}

\begin{figure}
    \centering
    \includegraphics[width=0.5\linewidth]{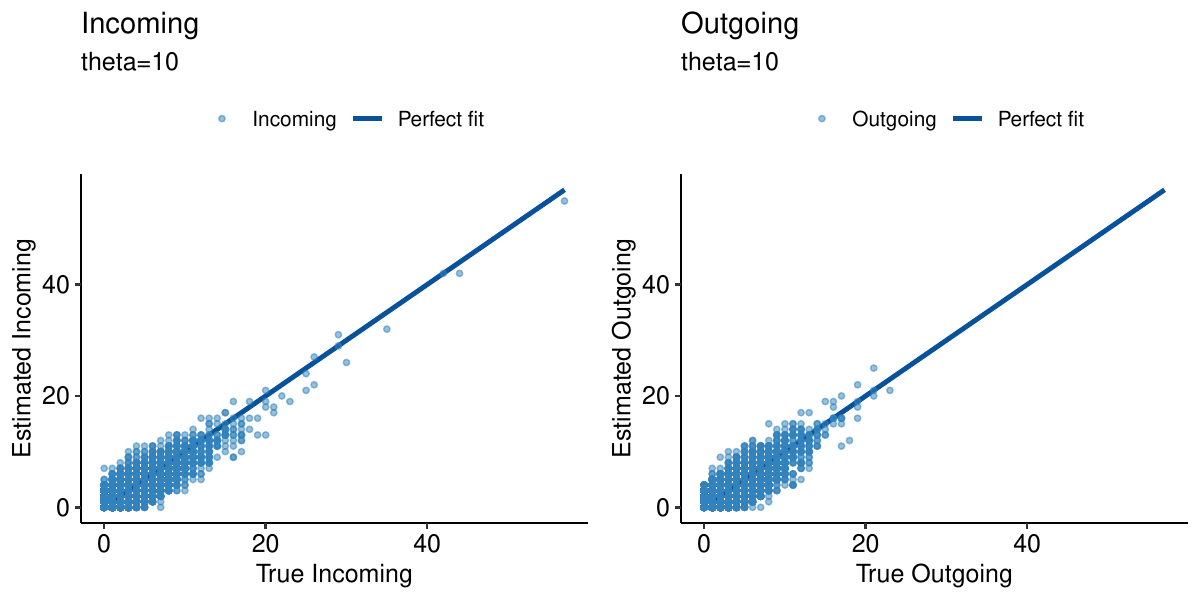}
    \caption{The estimated inflow and outflow for data with chosen data with Negative Binomial- $\theta=10$.}
    \label{fig:IncOut_Mis_10}
\end{figure}

Inspecting Figure \ref{fig:IncomOutSim}, and Figure \ref{fig:IncOut_Mis_05} to Figure \ref{fig:IncOut_Mis_10}, we observe an overestimation of inflow and outflow, which diminishes as the overdispersion reduces.


\end{document}